\documentclass[graybox]{svmult}

\usepackage{mathptmx}  
\usepackage{helvet}       
\usepackage{courier}     
\usepackage{type1cm} 
\usepackage{makeidx} 
\usepackage{graphicx}
\usepackage[caption=false]{subfig}
\usepackage{epstopdf}
\usepackage{multicol} 
\usepackage[bottom]{footmisc}
\usepackage{amsmath}
\usepackage{amssymb}
\usepackage{hyperref}

\makeindex

\begin{document}

\title*{The holographic Ricci dark energy and its possible doomsdays.}

\author{Moulay-Hicham Belkacemi, Mariam Bouhmadi-L\'{o}pez, Ahmed Errahmani and Taoufiq Ouali}

\institute{Moulay-Hicham Belkacemi \at Laboratory of Physics of Matter and Radiation, Mohammed I University, BP 717, Oujda, Morocco
\email{hicham.belkacemi@gmail.com}
\and Mariam Bouhmadi-L\'{o}pez \at Instituto de Estructura de la Materia, IEM-CSIC, Serrano 121, 28006 Madrid, Spain  \email{mariam.bouhmadi@iem.cfmac.csic.es}
\and Ahmed Errahmani \at  Laboratory of Physics of Matter and Radiation, Mohammed I University, BP 717, Oujda, Morocco \email{ahmederrahmani1@yahoo.fr}
\and  Taoufiq Ouali \at  Laboratory of Physics of Matter and Radiation, Mohammed I University, BP 717, Oujda, Morocco \email{ouali\_ta@yahoo.fr}}

\maketitle


\abstract*{It is well known that the holographic Ricci dark energy can induce some future doomsdays in the evolution of the universe. Here we analyse the possible avoidance of those doomsdays by invoking a modification to general relativity on the form of curvature effects.}

\abstract{It is well known that the holographic Ricci dark energy can induce some future doomsdays in the evolution of the universe. 
Here we analyse the possible avoidance of those doomsdays by invoking a modification to general relativity on the form of curvature effects.}


\section{Introduction}

A possible approach to explain the current acceleration of the universe is based on the holographic dark energy \cite{Li:2004rb,Hsu:2004ri}
. Such a phenomenological model is based on the idea that the energy density of a given system is bounded by a magnitude proportional to the inverse square of a lenght characterising the system \cite{Susskind:1994vu,Cohen:1998zx}. When this principle is applied to the universe as a whole, we  obtain the holographic dark enery \cite{Li:2004rb,Hsu:2004ri}. It turns out that there are many different ways of characterising the size of the universe and one of them is related to the inverse of the Ricci curvature of the universe. When the size of the universe is charaterised in such a way, we end up with the holographic Ricci dark energy (RDE) model \cite{Gao}, whose energy density reads: 
\begin{equation}
\rho _{\mathrm{H}}=3\beta M_{P}^{2}\left( \frac{1}{2}\frac{dH^{2}}{dx}%
+2H^{2}\right) ,  \label{HRDE}
\end{equation}%
where $M_{P}$ is the Planck mass, $x=-\ln (z+1)=\ln(a)$, $z$ is the redshift and $\beta$ is a
dimensionless parameter that measures the strength of the holographic component.  

A spatially flat Friedmann-Lema\^{\i}tre-Robertson-Walker (FLRW) universe filled with this kind of matter accelerates and therefore the RDE can play the role of dark energy on the Universe. It turns our that the asymptotic behaviour of the Universe depends crutially on the value acquired by $\beta$: (i) if $\beta \leq 1/2$ the universe is asymptotically de Sitter, otherwise (ii) the universe faces a big rip singularity \cite{bigrip} in its future evolution. 

Our main purpose in this paper is to see if we can appease the big rip appearing in some cases on the RDE by invoking some infra-red and ultra-violet curvature corrections. This two corrections can be quite important to remove the big rip singularity which takes place on the future and at high energy. The curvature corrections will be modeled within a 5-dimensional brane-world model with an induced gravity (IG) term on the brane and a Gauss-Bonnet term in the bulk \cite{brane}.

\section{The RDE model with curvature corrections}

We consider a DGP brane-world model, where the bulk action contains a GB
curvature term. The bulk corresponds to two symmetric pieces of a
5-dimensional (5d) Minkowski space-time. The brane is spatially flat and its
action contains an IG term. We assume that the brane is filled
with matter and RDE. Then, the modified Friedmann
equation reads \cite{brane}:

\begin{equation}
H^{2}=\frac{1}{3M_{P}^{2}}\rho +\frac{\epsilon }{r_{c}}\left( 1+\frac{%
8\alpha }{3}H^{2}\right) H,  \label{friedmann1}
\end{equation}%
where $H$ is the brane Hubble parameter, $\rho =\rho_{\mathrm{m}}+\rho_{%
\mathrm{H}}$ is the total cosmic fluid energy density of the brane which can
be described through a cold dark matter component (CDM) with energy density $%
\rho _{\mathrm{m}}$ and an holographic Ricci dark energy component with
energy density $\rho _{\mathrm{H}}$. The parameters $r_{c}$ and $\alpha$
correspond to the cross over scale and the GB parameter, respectively,
both of them being positive. The parameter $\epsilon$ in Eq.~(\ref{friedmann1})
can take two values: $\epsilon =1$, corresponding to the self-accelerating
branch in the absence of any kind of dark energy \cite{brane}; and $%
\epsilon =-1$, corresponding to the normal branch which requires a dark
energy component to accelerate at late-time (see for example 
\cite{preparation,BouhmadiLopez:2010pp,BouhmadiLopez:2008nf}).
For simplicity, we will keep the terminology: (i) self-accelerating branch when $\epsilon=1$ and
(ii) normal branch when $\epsilon=-1$.

The modified Friedmann equation (\ref{friedmann1}) can be rewritten as 
\begin{equation}
\dfrac{dE}{dx}=-\dfrac{\Omega _{m}e^{-3x}+\left( 2\beta -1\right)
E^{2}+2\epsilon \sqrt{\Omega _{r_c}}(1+\Omega _{\alpha }E^{2})E}{\beta E},
\label{variation of E}
\end{equation}
where $E(z)=H/H_{0}$ and
\begin{eqnarray}
\Omega _{m} &=&\frac{\rho _{m_{0}}}{3M_{P}^{2}H_{0}^{2}},\,\,\,\,\Omega
_{r_{c}}=\frac{1}{4r_{c}^{2}H_{0}^{2}},\text{\thinspace and }\,\Omega
_{\alpha }=\frac{8}{3}\alpha H_{0}^{2}\,  \notag \\
&&
\end{eqnarray}%
are the usual convenient dimensionless parameters and the subscripts $0$
denotes the present value (we will follow the same notation as in \cite%
{BouhmadiLopez:2008nf,preparation,Belkacemi:2011zk}). By evaluating the modified Friedmann equation 
at present and imposing that the brane is currently accelerating, we obtain a constraint on the parameter $\beta$ which depends 
on the chosen brane 
\begin{equation}
\left\{
\begin{array}{c}
\beta <\beta _{\mathrm{lim}}\text{ for \ \ }\epsilon =+1, \\
\beta >\beta _{\mathrm{lim}}\text{ \ for \ }\epsilon =-1,%
\end{array}%
\right.  \label{conditions for beta}
\end{equation}%
where
\begin{equation}
\beta _{\mathrm{lim}}=\frac{1-\Omega _{m}}{1-q_{0}}.  \label{betalim}
\end{equation}%
An estimation of $\beta _{\mathrm{lim}}$ can be obtained as follows: the brane would behave roughly (to be consistent with the present observations) as the $\Lambda$CDM  leading to $\beta _{\mathrm{lim}}\sim 0.43$.

Even though the modified Friedmann equation (\ref{variation of E}) cannot be solved analytically, we can obtain the future asymptotic behaviour of the brane which reads: (i) If $\beta < \beta_{\mathrm{lim}}$ or $\beta_- \leq \beta$, the brane is asymptotically de Sitter. (ii) If $\beta_{\mathrm{lim}}<\beta<\beta_-$, the brane faces a big freeze singularity in its future \cite{BouhmadiLopez:2006fu}, where (see also Fig. 1)
\begin{eqnarray}
\beta_\pm&=&\frac{1+\Omega_\alpha\pm2\sqrt{\Omega_\alpha}(1-\Omega_m)}{2%
\left[1+\Omega_\alpha\pm\sqrt{\Omega_\alpha}(1-q_0)\right]}.
\label{defbetapm}
\end{eqnarray}

\begin{figure}[t]
\begin{center}
\includegraphics[width=0.6\columnwidth]{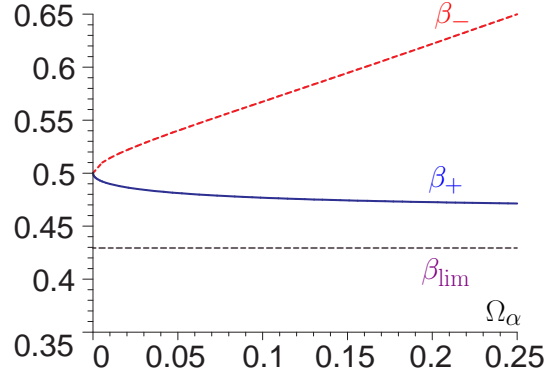}
\end{center}
\caption{Plot of the parameters $\protect\beta_\pm$ and $\protect\beta_{%
\mathrm{lim}}$, defined in Eqs.~(\protect\ref{defbetapm}) and (\protect\ref%
{betalim}), respectively, versus the parameter $\Omega_\protect\alpha$. We
have used the values $q_{0}\sim -0.7,$ and $\Omega
_{m}\sim 0.27$. The parameter $\protect\beta_{\mathrm{lim}}$ defines the
border line between the normal branch ($\protect\beta_{\mathrm{lim}}<\protect%
\beta$) and the self-accelerating branch ($\protect\beta<\protect\beta_{%
\mathrm{lim}}$).}
\label{plotbetas}
\end{figure}
We have completed and confirmed those results by solving numerically the cosmological evolution of the brane. We refer the reader to \cite {Belkacemi:2011zk} for more details.  Our analysis shows that even though the infra-red and ultra-violet effect can appease the big rip appearing on the RDE model, it cannot remove them completely. We would like as well to point out that when the GB term is switched off a little rip event \cite{Frampton:2011sp} can show up which is much milder that a big rip or a big freeze. The little rip has been previously found on brane-world model \cite{BouhmadiLopez:2005gk}.

\section{Conclusions}

We present an HRD energy brane-world model of the Dvali-Gabadadze-Porrati
scenario with a GB term in the bulk. The reason for invoking curvature corrections, for example  through a brane-world scenerio, is to try to smooth the doomdays present on a standard 4-dimensional HRD energy model. It turns out that the model presented here can only partially remove those doomsdays.


\begin{acknowledgement}
M.B.L. is supported by the Spanish Agency CSIC through JAEDOC064.
A.E. and T.O. are supported by CNRST, through the fellowship URAC 07/214410. This work was supported by the Portuguese Agency FCT through PTDC/FIS/111032/2009.
\end{acknowledgement}

\end{document}